\documentclass{aastex}
\usepackage{graphicx}
\usepackage{amsmath}
\usepackage{amssymb}
\usepackage{booktabs}
\usepackage{latexsym}
\usepackage{natbib}

\newcommand{\MSun}{M_\odot}
\newcommand{\Mmax}{m_{\mathrm{max}}}
\newcommand{\Mcl}{M_{\mathrm{cl}}}

\newcommand{\includeEPSx}[1]{\includegraphics*[scale=0.35, angle=270]{./#1}}
\newcommand{\includeEPS}[2]{\scalebox{#2}{\includegraphics*[angle=270]{./#1}}}
\newcommand{\customrule}{\cmidrule(r{.25em}){1-1}\cmidrule(lr{.25em}){2-4}\cmidrule(lr{.25em}){5-7}\cmidrule(l{.25em}){8-10}}
\newcommand{\customruletwo}{\cmidrule(r{.25em}){1-1}\cmidrule(lr{.25em}){2-4}\cmidrule(lr{.25em}){5-7}}

\title{First, second and third massive stars in Open Clusters}
\author{Alexey Mints}
\affil{University of Bielefeld, Germany}

\begin{document}

\begin{abstract}
 The goal of this paper is to study possibilities of using first, second and third massive stars 
 in open clusters to estimate total cluster mass and membership. We built estimator functions with the use of numerical simulations and analytical approximations and studied the precision and error distribution of the obtained estimator functions. We found that the distribution of the mass of first, second and third massive stars shows strong power-law
 tails at the high-mass end, thus it is better to use median or mode values instead of average ones.
 We show that the third massive star is a much better estimator then the first as it is more precise and less dependent on parameters such as maximum allowed stellar mass.
\end{abstract}

\maketitle

\section{Introduction}
Unfortunately, in most cases there does not exist a large body of statistics covering star cluster membership,
so estimation of the cluster mass and the full number of members is not an easy task. In some cases dynamical mass estimates were applied when spectroscopic data is available (making use of the virial theorem, see \citet{Kouwenhoven}), although this method can be imprecise (see \citet{Fleck}). 
Another method measures the total brightness of identified members and extrapolate it
using some luminosity function (see \citet{BB2005}). However it is often the case that only a few of the brightest
stars are reliably identified as members \citep{Kharchenko2003}, which provides only a tiny amount of information about the cluster.

It is natural to expect to find more massive stars in massive clusters. Although, assuming \citet{Salpeter} initial mass function (hereafter IMF) one would expect several stars with masses $M_{*} \sim 300 \MSun$ in our Galaxy, which is not the case. This controversy was discussed by \citet{Elmegreen} and he presented a relation between the
cluster mass and the mass of it's most massive star (assuming random sampling from the Salpeter IMF):

\begin{equation}
  \Mcl \sim 3 \times 10^3 \left( \frac{\Mmax}{100 \MSun} \right) ^{1.35} \MSun.
\end{equation}
Elmegreen tried to introduce an exponential cut-off function to explain the absence of heavy stars, while maintaining randomness of sampling from the IMF. However this led to a contradiction with the observed mass functions of massive clusters, as Salpeter-like power law can be traced up to at least $100 \MSun$. He proposed several explanations for this, including dependence of the IMF on the initial cluster mass.

Kroupa's IMF (\citet{Kroupa2001}) has become popular over the past decade as a standard cluster IMF.  
It is built from several power-law parts:
\begin{align} \label{eq:spectrum}
 f(m)dm &= C m^{-\alpha} dm  \nonumber \\
 F(m) &= \int_0^{m} f(m')dm'
\end{align} 
With parameters (\citet{Kroupa2001}):
\begin{align} \label{eq:kroupa} 
  \alpha_0 = +0.30 &\hspace{0.5cm} 0.01 \leq m/\MSun < 0.08, \nonumber \\
  \alpha_1 = +1.30 &\hspace{0.5cm}  0.08 \leq m/\MSun < 0.50, \\ 
  \alpha_2 = +2.35 &\hspace{0.5cm}  0.50 \leq m/\MSun < \Mmax. \nonumber 
\end{align} 
From Equation \ref{eq:kroupa} we see that a sharp cut-off at a high-mass end was introduced, although the exact value of $\Mmax$ is left as a free parameter.
Using this version of IMF, \citet{Origin} reviewed the correlation between the mass of the most massive star in the cluster and the total cluster mass. They arrived at the following conclusion: that random sampling contradicts observations (for a description of various samplings see Section \ref{sec:model}).
\citet{MassFromN} discussed correlation between the number of stars in the cluster and its most massive member mass, and found it compatible with random sampling. However, they only looked at a small range in cluster masses. These (and many more) papers were recently reviewed in \citet{Kroupa2010}.

\citet{Faustini} recently studied stellar clusters around young high-mass stars. They used Monte-Carlo simulations to 
try to reproduce properties of observed clusters. Among other results, they found that the distributions for the total cluster mass
and for the mass of the most massive star were skewed and suggested using the mode of the distribution instead of its mean or median values.

In this paper we try to produce cluster mass ($\Mcl$) and membership ($N$) estimators using masses of three most massive members,
and analyse the precision of these estimators. We concentrate on two questions: 
\begin{itemize}
 \item what data provides the most reliable information on the cluster properties?
 \item what is the best method to extract cluster properties from that data?
\end{itemize}
Having these two goals in mind means we will neglect at least three very important factors: stellar binarity, stellar evolution and cluster dynamics. \textit{Stellar binarity} can show itself directly by altering the IMF (if the process of binary star formation differs from that of a single star) or indirectly by unresolved binaries, that alter the luminosity function and our assumptions about the IMF. This problem was studied by \citet{WKM2008};
 \textit{Stellar evolution} is important for old clusters, as their heavy stars can evolve and turn to stellar remnants or lose a fraction of their initial mass, which is caused by stellar wind. Modelling this process requires a set of assumptions and is beyond the scope of this work. Note that \citet{Kroupa2010} studied young clusters which allowed them to neglect stellar evolution.
 \textit{Cluster dynamics} leads to mass segregation and evaporation of the cluster with time. Although it is believed that the lighter
stars are the most probable candidates to be ejected from the cluster, this can also happen to the heavier stars. For more 
details on this problem see \citet{Pflamm} and \citet{Fleck}. However, this effect is less important for young clusters.

Of course, it is not possible to weigh stars directly, but we can estimate their mass, mostly done by measuring their brightness. However, this process
is a very uncertain one. The other way to estimate stellar masses is to weigh binary members, if the orbit is known, but this raises problems of influence of the binarity
on the IMF mentioned above. But, again, as we are interested in the statistical side of the problem, we will postpone astrophysical
difficulties for later research. 

If we assume all stars in the cluster to be single, then the (initial) mass of each star depends only on 
the initial mass function. Here we will consider only one IMF --- from \citet{Kroupa2001} (see eq. \ref{eq:kroupa}). 
The value of $\Mmax$ is still a matter of debate, therefore several values will be considered in this paper ($50, 150$ and $300 \MSun$), with the main focus being put on $\Mmax = 150 \MSun$. Although $\Mmax = 50 \MSun$ is not a realistic value it is used here
to study the dependence of the results on $\Mmax$. We will try to see how $\Mmax$ influences the mass estimator precision.

The following notation will be used throughout the paper: $\bar{m}_1$, $\tilde{m}_1$ for the average and the median value of $m_1$ respectively (the same for $m_2$ and $m_3$). Another useful value is  the position of the peak of the $m_{1,2,3}$ distribution for a given $\Mcl$ or $N$ (mode of the distribution), which we designate as $\hat{m}_{1,2,3}$. Kroupa IMF (see Equation \ref{eq:kroupa}) has an average stellar mass $\bar{m} = 0.36 \MSun$ for $\Mmax = 150 \MSun$.

\section{Model}\label{sec:model}

Following \citet{Origin}, we used three different methods for generating cluster members:
\begin{description}
 \item[Random sampling] --- $N$ stars are taken randomly from the IMF, with $N$ ranging from 300 to 10000.
 \item[Constrained sampling] --- $\Mcl$ is fixed, then stars are taken from the IMF until their total mass surpass $\Mcl$. Thus some spread in $N$ is expected in this sample.
 \item[Sorted sampling] --- $\Mcl$ is also fixed, then $N' = \Mcl/\bar{m}$ stars are taken from the IMF. If $M' = \sum_{N'} m_i$ smaller than $\Mcl$, then $\Delta N = (\Mcl - M')/\bar{m}$ stars are added to the cluster, giving a new $N'$ and $M'$ and the procedure is repeated the cluster mass $M'$ surpasses $\Mcl$. After this the stellar masses are sorted. If $|M' - \Mcl|$ is larger then $\left|\Mcl - (M'-m_{1})\right|$ then the heaviest star is removed from the set. 
\end{description}

According to \citet{Origin}, random sampling is the least realistic model, but the easiest to be modeled and described analytically.

For each set of parameters ($\Mcl$ or $N$, sampling, $\Mmax$), 30000 clusters were simulated, and for each one five values were saved: cluster mass $\Mcl$,
number of stars in the cluster $N$, and masses of the three most massive stars of the cluster --- $m_1, m_2, m_3$.

The goal is to build a method to find $\Mcl$ and/or $N$, when $m_1, m_2$ and $m_3$ is known.
It seems natural to find functions $\Mcl(\bar{m}_{1,2,3})$, $\Mcl(\tilde{m}_{1,2,3})$ and $\Mcl(\hat{m}_{1,2,3})$ (as well as $N(\bar{m}_{1,2,3})$, $N(\tilde{m}_{1,2,3})$ and $N(\hat{m}_{1,2,3})$). From here on they will be called \textit{mass estimators} (ME): average ME, median ME and mode ME.

\section{Analytics}
The probability for the most massive star to have mass $m_1 \in (m, m+dm)$ can be written (\citet{OrderStat}) as the probability for a given star to have mass in $(m, m+dm)$ multiplied by the probability that all other stars have masses below $m$ and by the number of stars $N$ (because any star can be the most massive one):
\begin{eqnarray}
 P(m_1 \in (m, m+dm)) &=& N f(m) \left[ F(m) \right]^{N-1} \nonumber \\ 
    &=& N f(m) \left[ 1 - \int_m^{\Mmax} f(m')dm' \right]^{N-1} \label{eq:prob1}
\end{eqnarray} 
Of course, $m_1$ should be smaller then $\Mmax$, otherwise $P \equiv 0$.

We can confidently use part of the Kroupa IMF (see Equation \ref{eq:kroupa}) for $m > 0.5 \MSun$, as the most massive stars are usually much heavier than $0.5 \MSun$.
Substituting \ref{eq:spectrum} into \ref{eq:prob1} and integrating we get (for $a \neq 1$):
\begin{equation}
  P(m_1 \in (m, m+dm)) = N C m^{-\alpha} \left[ 1 - \frac{C}{1-\alpha} \left( \Mmax^{1-\alpha} - m^{1-\alpha} \right) \right]^{N-1} 
\end{equation} 
where $C$ is a normalisation constant. 

If $N$ is large, then we can use exponent instead of square brackets (and replace $N-1$ by $N$ for the sake of simplicity):
\begin{equation} \label{eq:p_1st}
 P(m_1 \in (m, m+dm)) \simeq N C m^{-\alpha} \exp \left( -\frac{N C}{1-\alpha} \left( \Mmax^{1-\alpha} - m^{1-\alpha} \right)\right)
\end{equation} 
The maximum of this distribution (or the mode of distribution) is located at the point
\begin{equation}
 \hat{m}_1 = \left(\frac{NC}{\alpha}\right)^{1/(\alpha-1)}.
\end{equation} 

For $\hat{m}_1 \geq \Mmax$ the maximum is obviously at the point $\hat{m}_1 = \Mmax$. 
This puts an upper limit on the cluster mass that can be estimated with this formula. This is $N \approx 26000$ and thus
$\Mcl = \bar{m} N = 9500 \MSun$. By inverting this equation we can get an estimate for $N$ and $\Mcl$ from $\hat{m}_1$:
\begin{eqnarray}
  N &=& \frac{\hat{m}_1^{\alpha-1} \alpha}{C}, \nonumber \\ 
  \Mcl &=& \frac{\bar{m} \hat{m}_1^{\alpha-1} \alpha}{C}. \label{eq:mhat_est}
\end{eqnarray} 

Note that $\Mmax$ is hidden within the constant $C$ in these equations, although the dependence is weak.

For the $n$'th massive star, if $n \ll N$ we can use the expression:
\begin{equation}\label{eq:p_nth}
 P(m_n \in (m, m+dm)) \simeq \left( 1 - F(m) \right)^{n-1} P(m_1 \in (m, m+dm))
\end{equation} 

Finding average and median values for the equation \ref{eq:p_1st} is not that easy.

Building analytical expressions for the other sampling methods is a much more complicated task and is not discussed here.

\section{Results}
\subsection{Random sampling}\label{sec:random}
The random sampling model has as a natural parameter, the number of stars in the cluster, $N$. Here $N$ ranges from 300 to 10000, with 30000 clusters being simulated for each value of $N$.

\begin{figure}
  \begin{center}
   \includegraphics*[scale=0.50, angle=270]{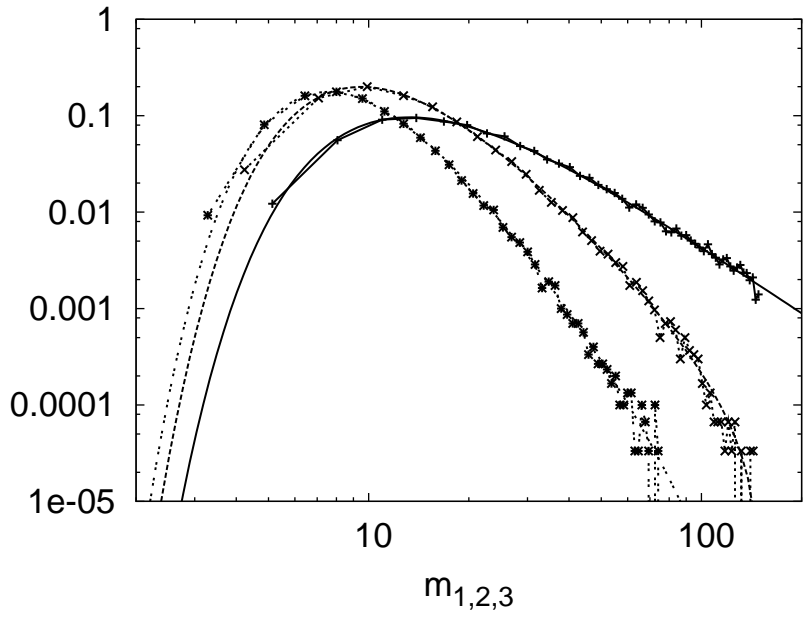}
  \end{center}
 \caption{Distribution of $m_1$ (plus signs), $m_2$ (crosses) and $m_3$ (stars) with theoretical estimates from Equations \ref{eq:p_1st} and \ref{eq:p_nth} (long-dashed, short-dashed and dot-dashed, respectively) for $N = 1000$ }  \label{fig:random_hist}
\end{figure}

For each value of $N$ the distributions of $m_1, m_2$ and $m_3$ were calculated. An example of these distributions is shown in Figure \ref{fig:random_hist}. From the figure it can be seen that the theoretical estimates given by Equations \ref{eq:p_1st} and \ref{eq:p_nth} match the data well.
Note the long power-law tails of the distributions, especially for $m_1$. This tail leads to significant differences between the average and the median values, making the average much higher. Thus, averages are not so well suited to making cluster mass estimators.

\begin{figure}
  \begin{center}
    \includeEPS{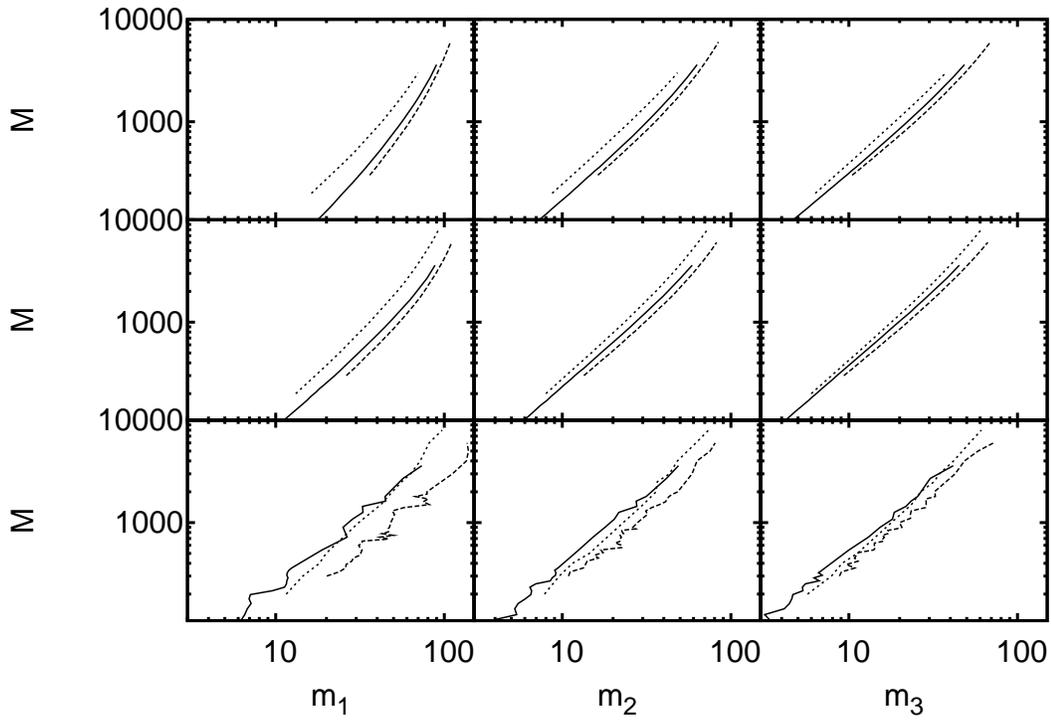}{0.45}
  \end{center}
  \caption{Estimators data: dependencies of $\Mcl$ on $m_1$ (left), $m_2$ (middle) and $m_3$ (right) for random (solid lines), constrained (long-dashed) and sorted (short-dashed) samplings. Top row is for average values, middle is for medians and bottom for the mode.}\label{fig:median1}
\end{figure}

Now the task is to build a method to find $\Mcl$ and/or $N$, knowing $m_1$ $m_2$ and $m_3$. We will try to find functions $\Mcl (\bar{m}_{1,2,3}), \Mcl( \tilde{m}_{1,2,3})$ and $Mcl ( \hat{m}_{1,2,3} )$. These functions for $\Mcl$ are shown in Figure \ref{fig:median1}. They can be approximated with functions of the shape:
\begin{eqnarray} \label{eq:fits}
  \Mcl(m_{1,2,3}) &=& a m_{1,2,3}^b (\Mmax-m_{1,2,3})^c \\ \nonumber
  N(m_{1,2,3}) &=& a m_{1,2,3}^b (\Mmax-m_{1,2,3})^c
\end{eqnarray} 
so those functions rise as power laws for small $m$ and then saturates as $m$ goes to $150 \MSun$ ($\Mmax$).

The parameters of the fits are shown in Table \ref{tbl:fits}. The first column refers to one of the functions from Equation \ref{eq:fits}, and the second, third and fourth columns are for different parameters applied to this function --- $a$, $b$ and $c$, respectively.
For $N(\hat{m}_{1,2,3})$ Eqautions \ref{eq:fits} can be used (by setting $c = 0$ and $b = 1.35$. For other estimators $b$ is always close to $\alpha_2 - 1 = 1.35$ and $c$ is close to $-1$, although $f(m) = a m^{1.35} (150-m)^{-1}$ is a bad fit. The value of $c$ decreases from $f(m_1)$ to $f(m_3)$ --- this is caused by the fact, that the values of $m_3$ are much smaller than $m_1$ and are well separated from $\Mmax$, therefore they are less affected by saturation. Thus, $f(m_3)$ is less sensitive to the value of $\Mmax$.
It should be also mentioned that $a, b$ and $c$ are highly correlated, so values given in the Table \ref{tbl:fits} might be not the only ones giving good fits.

\begin{table}
\begin{center}
\caption{Parameters of fits (see Eq. \ref{eq:fits})}\label{tbl:fits}
\begin{tabular}{cccccccccc} \\ \toprule
   \multicolumn{4}{c}{Fits based on... average values} 
 & \multicolumn{3}{c}{...median values}
 & \multicolumn{3}{c}{... mode values} 
 \\ \customrule
Function    &        a &   b  &     c &       a &    b &     c &      a &   b  &  c    \\ \customrule
$\Mcl(m_1)$ &   504.90 & 1.55 & -1.23 &  512.24 & 1.30 & -0.94 &  30.09 & 1.35 & 0.00  \\ \customrule
$\Mcl(m_2)$ &  2492.00 & 1.35 & -1.17 &  623.04 & 1.34 & -0.82 &  42.32 & 1.40 & 0.00  \\ \customrule
$\Mcl(m_3)$ &  4106.74 & 1.31 & -1.13 &  865.90 & 1.34 & -0.79 &  60.48 & 1.38 & 0.00  \\ \customrule
$N(m_1)$    &  1431.05 & 1.55 & -1.23 & 1894.15 & 1.26 & -0.97 &  83.58 & 1.35 & 0.00  \\ \customrule
$N(m_2)$    &  7091.91 & 1.35 & -1.17 & 2659.81 & 1.30 & -0.88 & 117.55 & 1.40 & 0.00  \\ \customrule
$N(m_3)$    & 11770.94 & 1.31 & -1.14 & 4241.68 & 1.31 & -0.89 & 168.00 & 1.38 & 0.00  \\ \bottomrule
\end{tabular}

\end{center}
\end{table}

\begin{figure} 
  \begin{center}
   \includeEPS{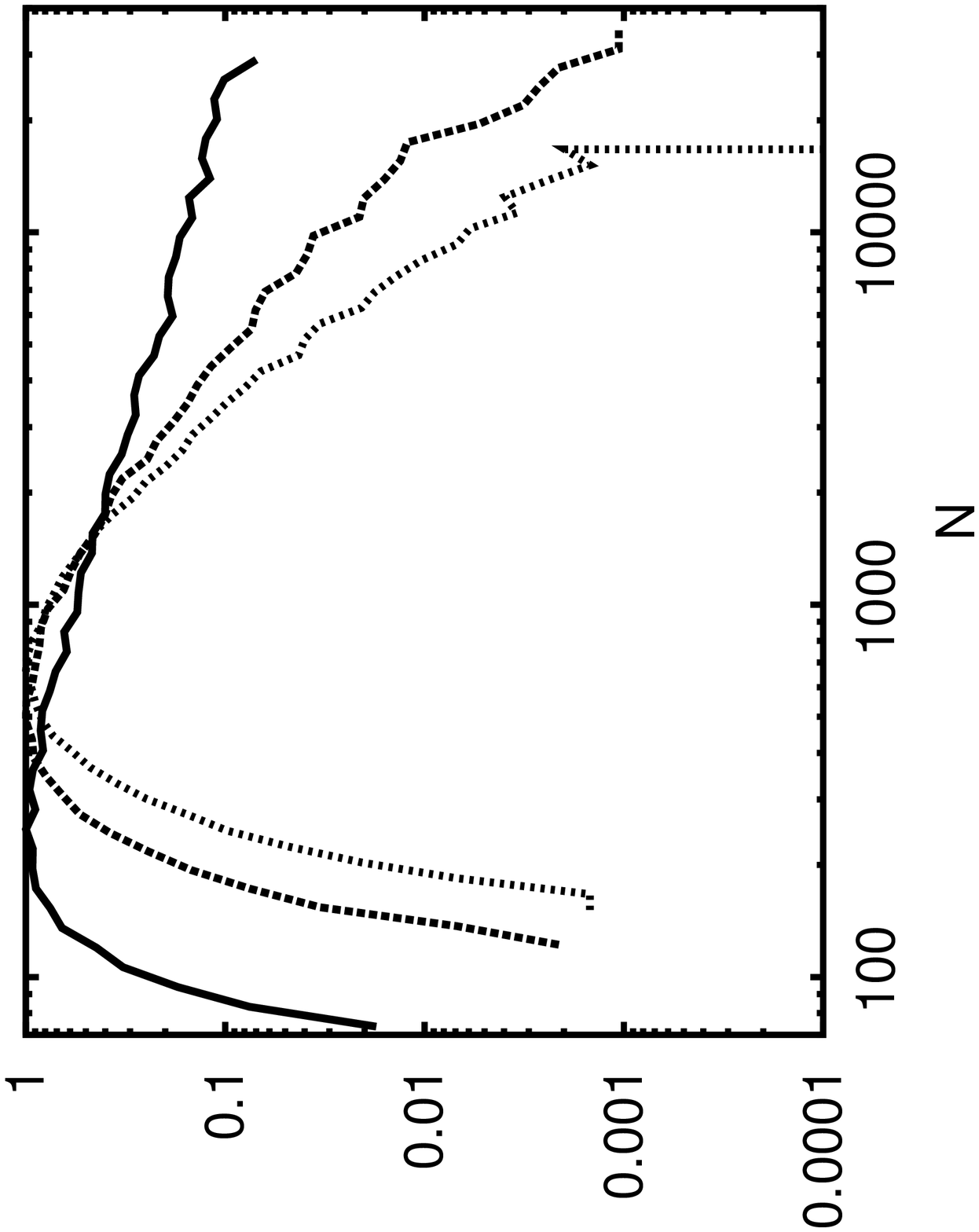}{0.3}
   \includeEPS{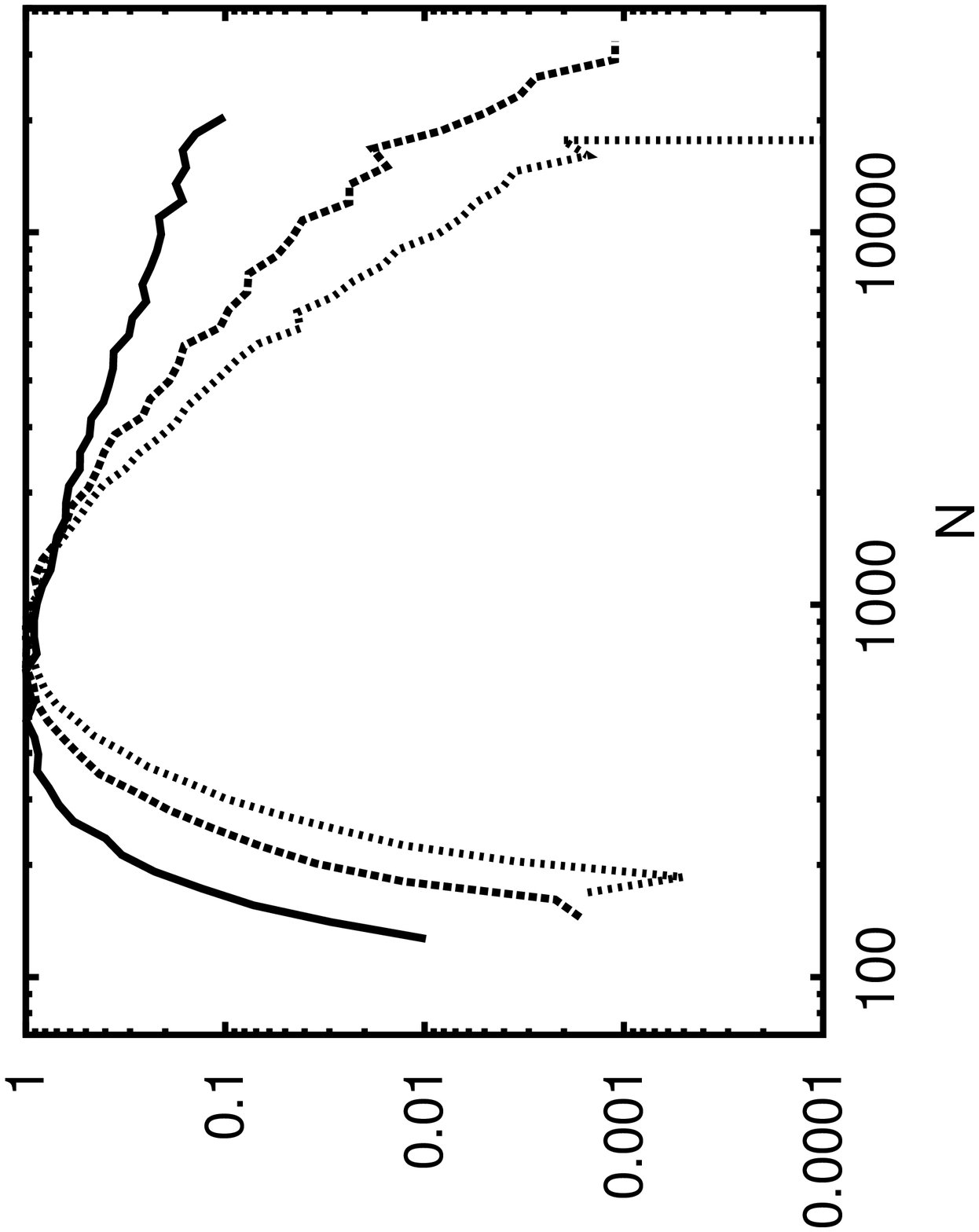}{0.3}
   \includeEPS{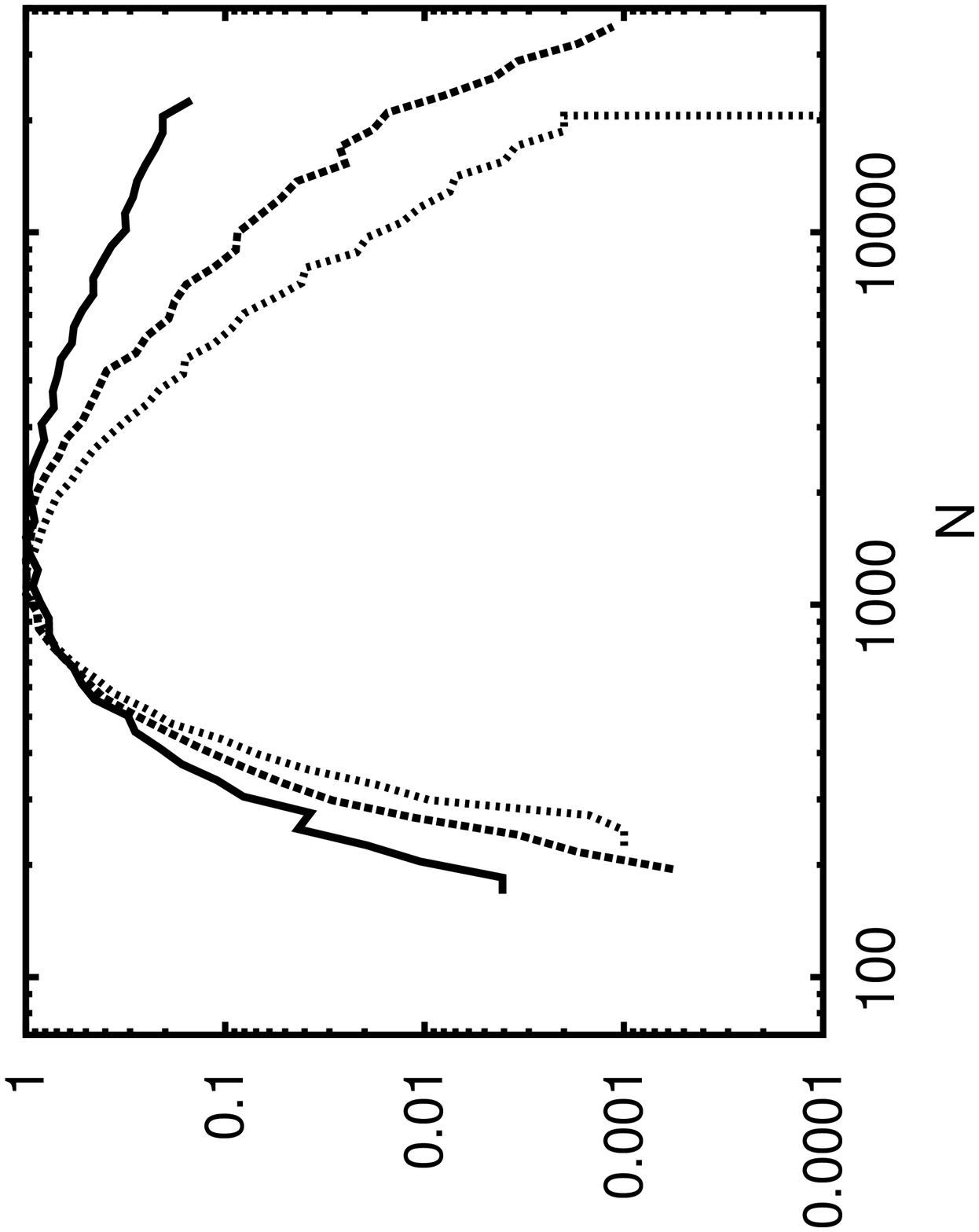}{0.3}
  \end{center}
 \caption{Distribution of estimates for a number of stars in the cluster $N(m_{1,2,3})$ (solid, long-dashed and short-dashed lines, respectively) for a cluster with pre-defined 1000 stars. Estimators are based on average (upper left), median (upper right) and mode (bottom) values. }\label{fig:errors}
\end{figure}

Given these approximations, we return to the initially simulated data to test how good they are. Namely, we will substitute $m_{1,2,3}$ for each cluster into mass estimators (see Eq. \ref{eq:fits}) to get $\Mcl(m_i)$ and $N(m_i)$, which can then be compared to the real values. This produces some distributions of estimated $\Mcl(m_i)$ and $N(m_i)$. Errors of the estimation can be calculated, as $|\Mcl(m_i) - \Mcl|$ and $|N(m_i)-N|$. A sample of the result for $N(m_i)$ is shown in Figure \ref{fig:errors} for a cluster with a pre-defined number of stars ($N=1000$).
Note that there is a large power-law tail at the high-mass side where $N$ goes to $10^6$, which is highest for $N(m_1)$ and smallest for $N(m_3)$ in all cases. Generally $N(m_3)$ shows a smaller spread than other estimators. The distribution of $N(m_3)$ also peaks closer to the real value $N=1000$. 

Tables \ref{tbl:estvalues} and \ref{tbl:estdisp} summarise the relative errors of mean and relative dispersions for various estimators 
and samplings. 
The average estimator is the worst one, giving the highest error of the average value in almost all cases --- sometimes up to 23\%,
with high dispersion.
The best one seems to be the median estimator, with errors of less than 2\%. 
The mode estimator is even worse than the average estimator for random sampling (75\% error), but it is better for the sorted one --- which is a more realistic sampling.
As expected, the result is due to the power-law tail of the distributions, to which the median (and mode) values are less sensitive. There is a high probability for $m_1$ to be close to $\Mmax$, where estimator functions (see Eq. \ref{eq:fits}) are very sensitive to $m_i$, thus producing a higher error and extremely large dispersions. The mode estimator is free from this effect by definition, as there is no $(\Mmax-m)^c$ factor.

The power-law tails of the distributions also cause extremely high dispersions for the estimates based on $m_1$. Dispersions (and errors of the mean) are much smaller for estimators based on $m_2$ and $m_3$, as the slope of the power-law tail is significantly higher. Relative dispersions are smallest for the mode estimator, but it has a higher relative error for the mean when compared with the median estimator.
Note, that in most cases estimators based on $m_3$ show the best results both in terms of the error of the mean and the dispersions.

Here we emphasise once again, that errors are distributed in a significantly non-Gaussian way in this problem. Using median values minimises the error for a high proportion of the data, while for a smaller proportion the errors remain large.

\begin{table} 
\begin{center}
\caption{Relative error (in percents) of mean value of estimated masses }\label{tbl:estvalues}
\begin{tabular}{cccccccccc} \\ \toprule
 & \multicolumn{3}{c}{Random sampling} &  \multicolumn{3}{c}{Constrained sampling} & \multicolumn{3}{c}{Sorted sampling} \\ \customrule
Value      & $f(m_1)$ & $f(m_2)$ & $f(m_3)$ & $f(m_1)$ & $f(m_2)$ & $f(m_3)$& $f(m_1)$ & $f(m_2)$ & $f(m_3)$\\ \customrule
 & \multicolumn{9}{c}{Median estimator} \\ \customrule
 N &  0.88 &  0.28 &  0.41 &  0.48 &  0.32 &  0.27 &  1.77 &  1.89 & 1.90  \\ \customrule
 M &  1.37 &  0.99 &  1.05 &  0.36 &  0.20 &  0.12 &  0.36 &  0.23 & 0.13  \\ \customrule
 & \multicolumn{9}{c}{Average estimator} \\ \customrule
 N & 23.23 & 20.13 & 15.29 & 12.44 & 13.66 & 11.43 & 18.09 & 10.41 & 6.43  \\ \customrule
 M & 23.24 & 20.14 & 15.29 & 12.48 & 13.69 & 11.45 & 19.08 & 11.99 & 8.44  \\ \customrule
 & \multicolumn{9}{c}{Mode estimator} \\ \customrule
 N & 74.97 & 43.85 & 25.87 & 19.44 & 15.66 & 12.99 &  7.41 &  6.35 & 4.17  \\ \customrule
 M & 74.82 & 43.73 & 25.76 & 21.27 &  9.29 &  7.79 &  7.92 &  7.63 & 4.67  \\ \bottomrule
\end{tabular}
\end{center}
\end{table}

\begin{table} 
\begin{center}
\caption{Relative dispersion (in percents) for mass estimations}\label{tbl:estdisp}
\begin{tabular}{cccccccccc} \\ \toprule
 & \multicolumn{3}{c}{Random sampling} & \multicolumn{3}{c}{Constrained sampling} & \multicolumn{3}{c}{Sorted sampling} \\ \customrule
Value      & $f(m_1)$ & $f(m_2)$ & $f(m_3)$ & $f(m_1)$ & $f(m_2)$ & $f(m_3)$& $f(m_1)$ & $f(m_2)$ & $f(m_3)$\\ \customrule
 & \multicolumn{9}{c}{Median estimator} \\ \customrule
 N &  46.02 & 1.80 & 0.74 & 13.56 & 1.84 & 0.66 &  122.50 & 1.17 & 0.43  \\ \customrule
 M &  39.00 & 1.62 & 0.70 & 12.97 & 1.77 & 0.65 &  111.41 & 1.13 & 0.43  \\ \customrule
 & \multicolumn{9}{c}{Average estimator} \\ \customrule
 N & 264.68 & 3.38 & 0.82 & 35.47 & 3.11 & 0.73 & 1296.57 & 1.99 & 0.46  \\ \customrule
 M & 259.82 & 3.34 & 0.81 & 34.34 & 3.04 & 0.73 &  547.78 & 1.00 & 0.36  \\ \customrule
 & \multicolumn{9}{c}{Mode estimator} \\ \customrule
 N &   1.08 & 0.84 & 0.58 &  0.41 & 0.60 & 0.42 &    0.77 & 0.44 & 0.32  \\ \customrule
 M &   1.08 & 0.84 & 0.58 &  0.38 & 0.56 & 0.39 &    0.79 & 0.46 & 0.33  \\ \bottomrule
\end{tabular}
\end{center}
\end{table}

\subsection{Constrained sampling}
We applied almost the same algorithm, as in the random sampling case, for the constrained sampling case. The only change was that we did not have an analytical formula for $\hat{m}_{1,2,3}$, and therefore had to use fits to the simulated data of the shape $f(\hat{m}) = a \hat{m}^b$.

\begin{figure}
  \begin{center}
   \includeEPSx{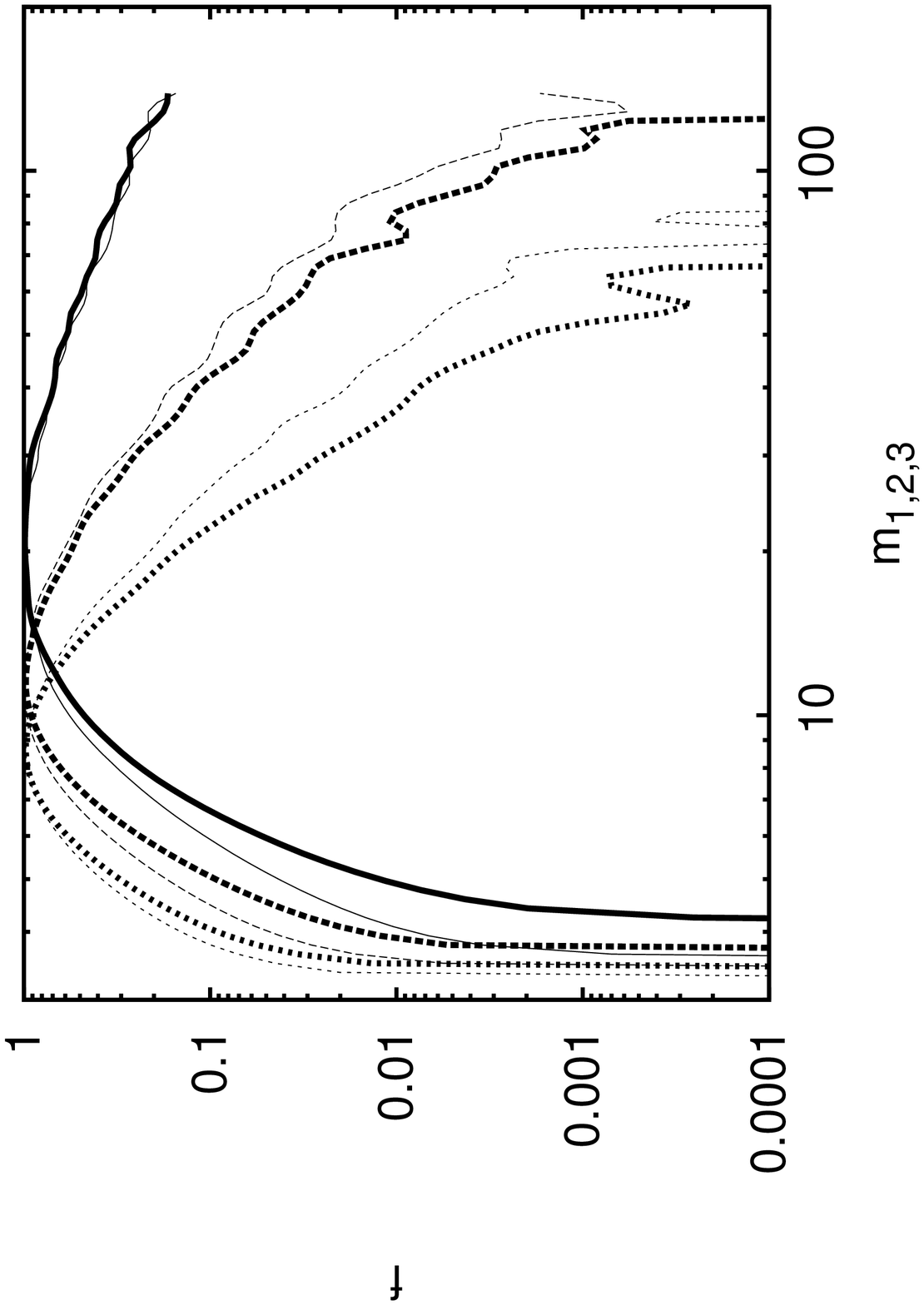}
   \includeEPSx{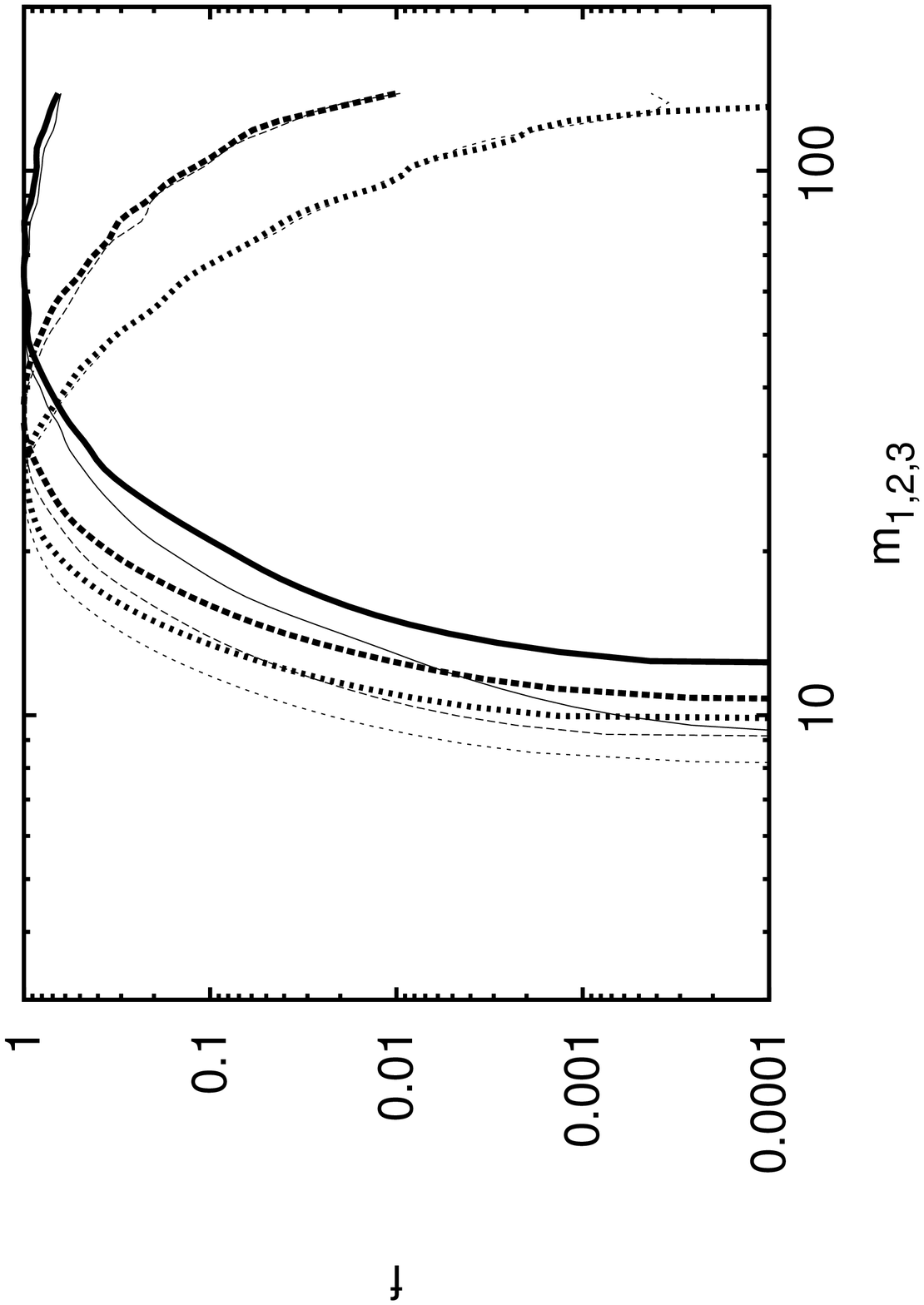}
  \end{center}
 \caption{Normalised distribution of $m_1$ (solid), $m_2$ (long-dashed) and $m_3$ (short-dashed) for constrained sampling (thick lines) and for random sampling (thin lines, same as Figure \ref{fig:errors}). Top panel: $N = 1000; \Mcl \approx 300$. Bottom panel:  $N = 4500; \Mcl \approx 1500$.  } \label{fig:compare2}
\end{figure}

In Figure \ref{fig:compare2} one can see that the difference between random and constrained samplings is not very large in most cases. The distribution for constrained sampling rises and falls faster than the one for random sampling. The faster decrease at the distribution high end for the small cluster (Figure \ref{fig:compare2}, top panel) is due to the fact that during the simulation the total mass comes close to the desired $\Mcl$, massive stars are preferentially rejected from the sample, when adding them will make the cluster too massive. Obviously, this effect vanishes for higher $\Mcl$, as one can see from the bottom panel in Figure \ref{fig:compare2}.

\subsection{Sorted sampling}\label{sec:sorted}

\begin{figure} 
  \begin{center}
   \includeEPSx{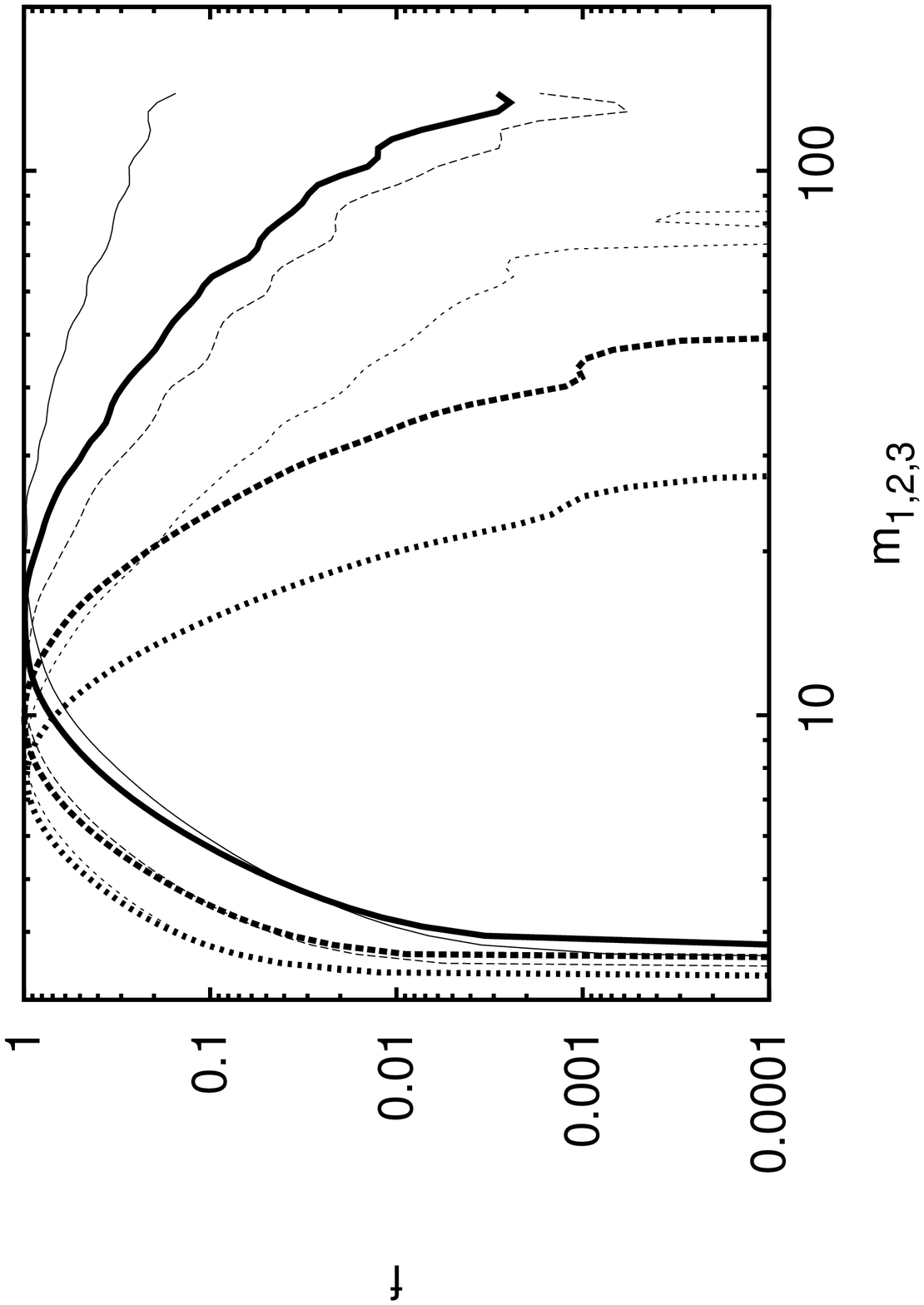}
   \includeEPSx{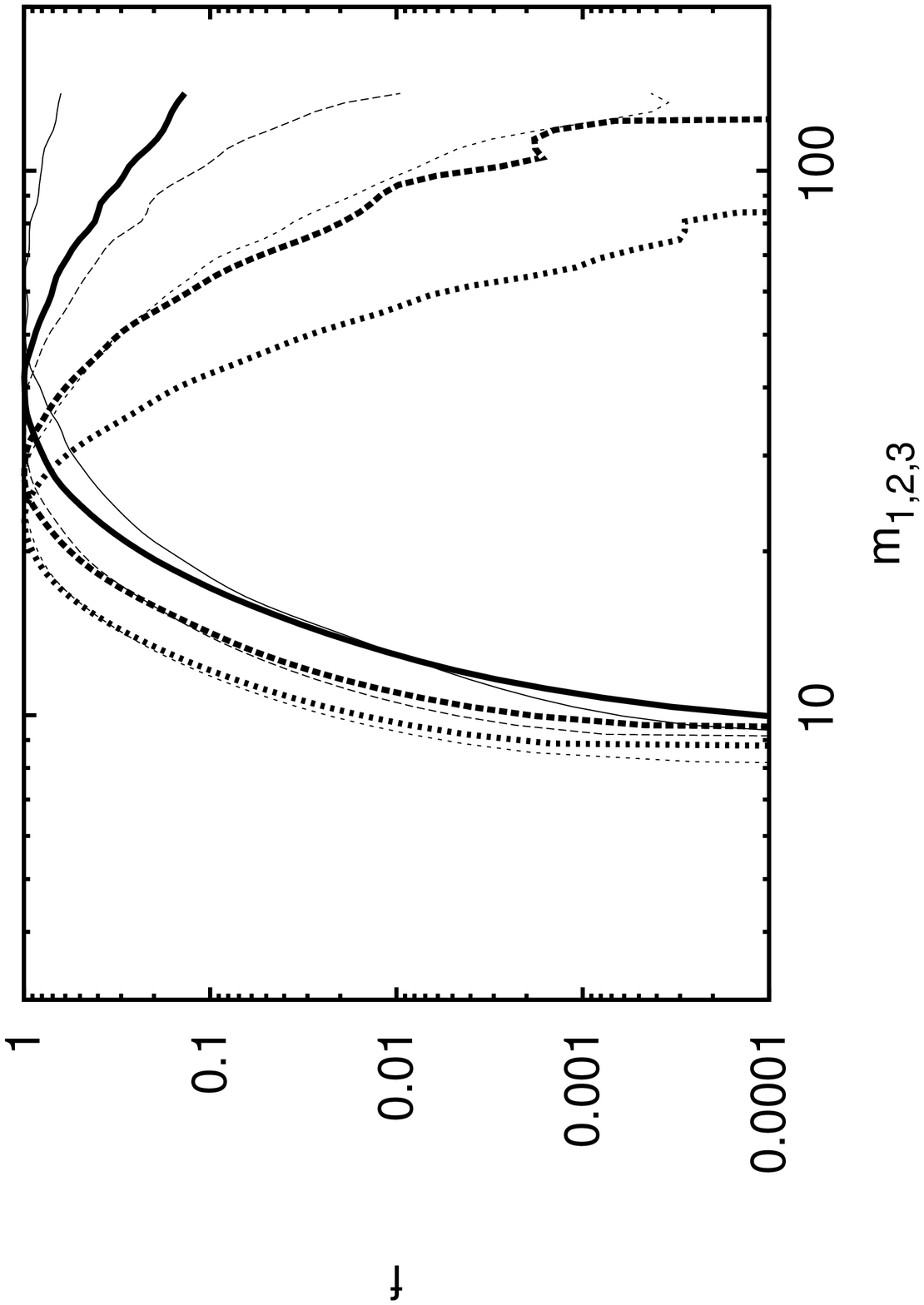}
  \end{center}
 \caption{Normalised distribution of $m_1$ (solid), $m_2$ (long-dashed) and $m_3$ (short-dashed) for sorted sampling (thick lines) and for random sampling (thin lines, same as at figure \ref{fig:errors}). Top panel: $N = 1000; \Mcl \approx 300$. Bottom panel:  $N = 4500; \Mcl \approx 1500$.  }\label{fig:compare3}
\end{figure}

Sorted sampling should suppress the probability of high-mass star formation even more than  constrained sampling. This can be seen on Figure \ref{fig:compare3}: distribution of $m_1$ for sorted sampling is almost like that for $m_2$ for random sampling at the high-end.

\subsection{Estimator reliability}

We first attempt to compare the various estimators by comparing the data on which they are constructed. Let us return to Figure \ref{fig:median1}. It is obvious that the curves are very close to each other. This can also be seen from the similarities  of fits parameters $a, b, c$ (see Eq. \ref{eq:fits} and Table \ref{tbl:fits}). So one might expect that predictions made with different estimators for the same value of $m$ will not differ from each other significantly. This difference can be even smaller than the difference between the estimated and real value, as predictions can deviate from the real value in the same direction.
We calculate average difference as:
\begin{equation}
  \Delta f(m_i) = 100\%*\left< \left| \frac{f(m_i) - f_\mathrm{random}(m_i)}{f_\mathrm{random}(m_i)}\right| \right>, \label{eq:difference}
\end{equation} 
where $i = 1,2,3$ and $m_i$ goes from $3$ to $140 \MSun$. This is a measure of how far away the estimators are from each other on average. The result is shown in Table \ref{tbl:variance}. Note that in most cases $\Delta f(m_3) < \Delta f(m_1)$. 
Constrained sampling is much closer to random sampling than to sorted sampling (from 14 to 71\%, comparing to 21--92\%). 
The reason for this is, of course, that different samplings give different $m_i$ distributions that are used to produce estimators. This can be seen in Figure \ref{fig:median1} by the distance between the lines. 
Differences remain large, on the order of 20\%, which is much larger than the relative errors of the mean value (see Table \ref{tbl:estvalues}) and relative dispersions (see Table \ref{tbl:estdisp}). This difference is less important for estimators based on $m_1$, as the relative errors of the mean value and relative dispersions are comparable to the differences between samplings. 
Due to this fact it is not efficient to use statistics on the most massive stars for distinguishing between various samplings.

Thus it is crucial to know which sampling method is more realistic, although there is still some discussion about it (see the Introduction).
It is also important to notice, that current mass estimates for both $\Mmax$ and $\Mcl$ can have errors as high as 50\%.

\begin{table}
\begin{center}
\caption{Relative difference (in percents, see Eq. \ref{eq:difference}) between estimators for random and other sampling} \label{tbl:variance}
\begin{tabular}{cccc} \\ \toprule
Sampling & $\Delta f(m_1)$ & $\Delta f(m_2)$ & $\Delta f(m_3)$ \\ \midrule
\multicolumn{4}{c}{Average estimator} \\ \midrule
Constrained & 24 & 17 & 18  \\ \midrule
Ordered     & 92 & 38 & 21  \\ \midrule
\multicolumn{4}{c}{Median estimator} \\ \midrule
Constrained & 20 & 14 & 14  \\ \midrule
Ordered     & 89 & 55 & 45  \\ \midrule
\multicolumn{4}{c}{Mode estimator} \\ \midrule
Constrained & 71 & 66 & 67  \\ \midrule
Ordered     & 34 & 45 & 47  \\ \bottomrule
\end{tabular}
\end{center}
\end{table}

\begin{table}
\begin{center}
\caption{Relative error for estimators applied to the dataset different to the one the functions were built on (Sorted sampling only).} \label{tbl:wrong_data}
\begin{tabular}{ccccccc} \\ \toprule
$\Mmax$ (source dataset) & \multicolumn{6}{c}{$\Mmax$ (target dataset)} \\ \customruletwo
 & \multicolumn{3}{c}{Average ME} & \multicolumn{3}{c}{Median ME} \\ \customruletwo
 & 300 & 150 & 50 & 300 & 150 & 50 \\ \customruletwo
 & \multicolumn{6}{c}{Estimators based on $m_1$} \\ \customruletwo
 300 & 26.37 & 51.03 & 81.12 & 2.10 & 34.47 & 74.66  \\ \customruletwo
 150 &     - & 18.09 & 77.91 &     - & 1.77 & 70.67  \\ \customruletwo
  50 &     - &     - & 16.53 &     - &     - & 2.36 \\ \customruletwo
 & \multicolumn{6}{c}{Estimators based on $m_3$} \\ \customruletwo
 300 & 7.28 & 21.97 & 56.21 & 2.12 & 17.44 & 53.82  \\ \customruletwo
 150 &    - &  6.43 & 51.76 &    - & 1 .90 & 49.36  \\ \customruletwo
  50 &    - &    -  & 5.43  &    - &     - &  2.30 \\ \bottomrule
\end{tabular}
\end{center}
\end{table}

Another check for the reliability of the obtained estimators is to try to apply them to the ``wrong'' dataset, for example --- using the median estimator from sorted sampling (see Section \ref{sec:sorted}) to estimate masses for the random one (see Section \ref{sec:random}) or to use an estimator from a dataset with $\Mmax = 150 \MSun$ to estimate masses for the $\Mmax = 300 \MSun$ dataset etc.
An example of this is shown in Table \ref{tbl:wrong_data}. Here $\Mmax$ was varied: a dataset with one $\Mmax$ was used to build an estimator function (source dataset) that was then applied to the dataset with another $\Mmax$ (target dataset). It should be noted, that 
$\Mmax$ in the target dataset cannot exceed that of the source dataset, as the function from Equation \ref{eq:fits} will be undefined.
Diagonals in this Table (i.e., values with equal source and target datasets) are the same as columns 8 and 10 in Table \ref{tbl:estvalues}. As expected, errors increase with increasing difference between the source and target dataset's $\Mmax$.
However $m_3$ is better in almost all cases, but the median estimator is as sensitive to $\Mmax$ variations as the average one, showing
errors of up to 75\%. On the other hand, relative estimate dispersions decrease rapidly with $\Mmax$ difference, as the probability to
have $m_1$ close to $\Mmax$, where one can get large errors, becomes smaller. It is also important that values of $\Mmax$ that are smaller than $150 \MSun$ are not realistic, and $\Mmax = 50 \MSun$ was introduced just to study the effect of a large range of parameter values. 

\section{Conclusions}

Several mass estimators for cluster mass from the first, second and third most massive stars were defined in this paper. Their precision was estimated. Estimators based on the mass of the third massive member $m_3$ 
gave the best results (approximately 3-5 times better than those based on $m_1$), and are less dependent on the maximum allowed stellar mass $\Mmax$ and assumed way of star formation (algorithm for picking masses from the IMF). 
We found that it is also better to build estimators on the median or mode values of $m_i$ instead of the average values. The reason is that the strong power-law tails in the  $m_i$ distributions make the average value a less representative parameter.

The most important parameter is the assumed algorithm describing how the cluster mass is distributed among stars.

However, as several astrophysical effects were not taken into account, these results cannot yet be applied to most of the real clusters. Inclusion
of evolution into this model is a subject for further work. Here it was shown that $m_3$ is a good candidate for building 
mass estimators. Error analysis was also carried out and revealed a power-law tail in the error distribution. We showed that the median (or mode) values 
are much better sources for mass estimators than the average values.


\end{document}